\documentclass[11pt,english]{iopart}   
\usepackage{cite}
\usepackage{graphicx}  
\usepackage{dcolumn}  
\usepackage{bm}       
\usepackage[thinlines,thiklines]{easybmat}
\usepackage[T1]{fontenc}
\usepackage{bbold}

\usepackage{dsfont}
\usepackage{setspace}
\usepackage{subfigure}

\usepackage{iopams} 
\newcommand{\binom}[2]{{#1 \choose #2}}

\newcommand{\WW}{\mathcal{W}}

\newcommand{\bra}[1]{\ensuremath{\langle#1|}}

\newcommand{\ket}[1]{\ensuremath{|#1\rangle}}

\newcommand{\braket}[2]{\ensuremath{\langle #1|#2\rangle}}

\newcommand{\ketbra}[1]{\ensuremath{| #1 \rangle \langle #1 |}}
\newcommand{\kommentar}[1]{}

\newcommand{\be}{\begin{equation}}
\newcommand{\ee}{\end{equation}}
\newcommand{\eea}{\end{eqnarray}}
\newcommand{\bea}{\begin{eqnarray}}

\begin{document}

\title{Entanglement criteria for Dicke states}
\author{Marcel Bergmann$^{1,2}$ and Otfried G\"uhne$^2$}
\address{$^1$ Institut f\"ur Physik, Johannes-Gutenberg-Universit\"at, 
Staudingerweg 7, D-55128 Mainz, Germany}
\address{$^2$ Naturwissenschaftlich-Technische Fakult\"at, 
Universit\"at Siegen, Walter-Flex-Str. 3, D-57068 Siegen, Germany}

\date{\today}

\begin{abstract}
Dicke states are a family of multi-qubit quantum states with 
interesting entanglement properties and they have been observed in 
many experiments. We construct entanglement witnesses 
for detecting genuine multiparticle entanglement in the vicinity 
of these states. We use the approach of PPT mixtures to derive 
the conditions analytically. For nearly all cases, our criteria 
are stronger than all conditions previously known.
\end{abstract}


\section{Introduction}
Multiparticle entanglement plays a crucial role for different aspects 
of quantum information processing. Therefore, many researchers work 
on the theoretical characterization of entanglement or the experimental 
observation of entangled states using ions, photons, or solid-state 
systems \cite{hororeview, gtreview}. Among the different types of multiparticle quantum states, 
Dicke states have attracted a lot of attention. These states were first 
investigated in 1954 by R. Dicke for describing light emission from a 
cloud of atoms \cite{dicke54}, and recently several other features have been 
studied: Dicke states are relatively robust to decoherence \cite{bodoky}, 
their permutational symmetry allows to simplify the task of state tomography 
\cite{gezapi,  tobipi} and entanglement characterization \cite{gezajosa, gezanjp, 
huber, novo}. In addition, they are the symmetric states which are in some sense 
far away from the separable states \cite{dickedefinetti}. Finally, they are 
relatively easy to generate in photon experiments, and Dicke states with up 
to six photons have been observed experimentally \cite{hartmut, dicke4, dicke6}.

In this paper we develop an analytical approach to characterize genuine 
multiparticle entanglement in Dicke states. Our methods use the idea of
PPT mixtures \cite{bastian}. This means that the set of all biseparable states is 
approximated by the set of all states that can be written as a mixture of 
states which have a positive partial transpose (PPT) with respect to some 
partition. This approach has turned out to be fruitful for characterizing 
genuine multiparticle entanglement: For many examples, it delivers the 
strongest entanglement criteria so far \cite{bastian}, and it can be shown 
that it solves the problem of entanglement characterization for many 
families of states, such as permutationally invariant three-qubit states 
\cite{novo} or graph-diagonal four-qubit states \cite{gjmw}. For the family of 
permutationally invariant states (to which the Dicke states belong) 
it was shown in Ref.~\cite{novo} that one can decide the question whether a 
given state is a PPT mixture or not numerically with an effort scaling 
polynomially with the number of particles. In practice, however, the 
numerical approach is limited to systems with up to ten qubits, 
and therefore analytical results for arbitrary qubit numbers are highly 
desirable.
 
This paper is structured as follows. In Section 2 we explain the main
definitions and concepts which we need for our presentation. This includes 
the concept of genuine multiparticle entanglement and the notions  of 
entanglement witnesses, PPT-mixtures and the definition of Dicke states.
In Section 3 we determine the projective entanglement witnesses for Dicke
states with an arbitrary number of excitations. This completes existing work
where these witnesses were derived two special cases \cite{gezajosa, hartmut}; 
moreover, 
these witnesses will be a starting point for the further improvement 
with the PPT mixture approach. 
In Section 4 we first improve the projective entanglement witness for Dicke 
states with $N/2$ excitations by a more advanced ansatz and then generalize 
this to the case of Dicke states with arbitrary excitations. This section
contains the main results of this paper. We also compare the resulting 
noise tolerances with the best known results from the literature.
Finally, Section 5 is concerned with entanglement witnesses for W states, 
these are the Dicke states with only one excitation. Besides the characterization
via PPT mixtures, we present a different construction which turns out to be
stronger for this special case. Finally, we conclude and discuss possible 
generalizations of our results. 

\section{Basic concepts}
\subsection{Multipartite entanglement and entanglement witnesses}

We begin by explaining the structure of the set of entangled states 
in multiparticle systems. For reasons of simplification, all definitions 
are given for three-particle states, but it should be stressed that they 
can be extended to an arbitrary number of particles in a 
straightforward manner.

For three or more particles, different notions of entanglement 
exist\footnote{A detailed discussion of the various definitions
can be found in Ref.~\cite{gtreview}.}, 
but the most investigated class of entanglement is genuine multiparticle
entanglement \cite{hororeview, gtreview}. First, a three-particle state is called biseparable if 
it can be written as a convex combination of states which are separable 
for some bipartition,
\begin{equation}
\varrho^{bs}=p_{1}\varrho^{\rm{sep}}_{A|BC}+p_{2}\varrho^{\rm{sep}}_{C|AB}+p_{3}\varrho^{\rm{sep}}_{B|AC}.
\end{equation}
That means that the set of biseparable states is given by the convex hull 
of the states that are separable for a fixed bipartition. Any state that 
is not a biseparable state is called genuine multiparticle entangled. 
For more than three particles, this definition can directly be extended, 
only the number of possible bipartitions increases. To prove that a given 
state is entangled, one has to check whether it can be written in the 
above form or not. Unfortunately, it is not efficiently feasible to 
find all possible decompositions of the desired form.

An efficient tool for the detection of genuine multipartite entanglement 
which is frequently used in experiments are entanglement witnesses \cite{gtreview}. 
A witness operator is defined as an observable which has a positive 
expectation value on all biseparable states, but a negative expectation 
value on at least one genuine multiparticle entangled state. Experimentally, 
entanglement can then be proven by measuring a negative expectation value. 
Compared to other detection tools, e.g. Bell inequalities, the use of 
entanglement witnesses is advantageous because for every entangled state 
there exists an entanglement witness that detects it. 

The fact that every entangled state is detected by at least one 
entanglement witness leads to the question of how to find an suitable 
witness for a given quantum state. A typical construction goes as follows: 
for every pure state $|\psi\rangle$, the operator
\begin{equation}
\WW=\alpha\mathds{1}-|\psi\rangle\langle\psi|
\end{equation}
is called the projective entanglement witness \cite{gtreview}. Here $\alpha$ is 
the squared maximal overlap between $|\psi\rangle$ and the biseparable 
states, $\alpha= \sup_{\ket{\phi} \in {\rm bisep}}|\braket{\psi}{\phi}|^2.$
The maximal overlap $\alpha$ can be computed explicitly by taking the 
square of the maximal Schmidt coefficient, in addition maximized over 
all possible bipartitions.

\subsection{Dicke states}
The focus of this article is on entanglement witnesses 
for Dicke states. An $N$-qubit Dicke state with $k$ excitations
is defined as \cite{dicke54, gezajosa} 
\begin{equation}
|D^{N}_{k}\rangle = \frac{1}{\sqrt{\binom{N}{k}}}
\sum\limits_{j}P_{j}\big\{|1\rangle^{\otimes k}\otimes|0\rangle^{\otimes (N-k)}\big\},
\end{equation}
where $\sum_{j}P_{j}\{\cdot\}$ denotes the sum over all possible 
permutations. For example,
$
|D^{3}_{2}\rangle=\frac{1}{\sqrt{3}}(|110\rangle+|101\rangle+|011\rangle).
$
The case of  $k={N}/{2}$ excitations for even $N$
is the most discussed example of a Dicke state. In this case,  an 
analytical expression of the projective entanglement witness was derived
\cite{gezajosa}
\begin{equation}
\WW=\frac{1}{2}\frac{N}{N-1}\mathds{1}-|D^{N}_{\frac{N}{2}}\rangle\langle D^{N}_{\frac{N}{2}}|.
\label{gezaproj}
\end{equation}
Another important case is the Dicke state with $k=1$ excitations, 
these states are also called W states \cite{dur00},
\begin{equation}
|W_{N}\rangle=\frac{1}{\sqrt{N}}(|100\dots0\rangle+|010\dots0\rangle+\dots+|000\dots1\rangle).
\end{equation}
In this case, the projective witness is known to be \cite{hartmut}
\begin{equation}
\WW=\frac{N-1}{N}\mathds{1}-|W_{N}\rangle\langle W_{N}|.
\end{equation}
We stress that other types of witnesses and entanglement criteria for
Dicke states are known \cite{huber}, later we will also compare them with our 
results.

\subsection{The method of PPT mixtures}

\begin{figure}[t]
\centering
 \includegraphics[width=0.5\textwidth]{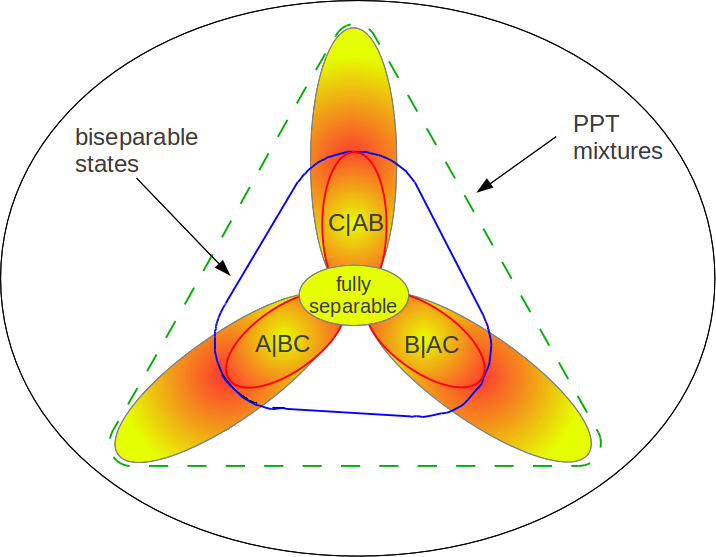}
\caption{Schematic view of the PPT mixtures (green, dashed lines)
as an outer approximation of the set of biseparable states (blue, 
solid lines). See the text for further details.}
\end{figure}

In order to improve the existing entanglement criteria for Dicke 
states, we consider a more general type of witnesses than the 
projective witnesses, the so-called fully decomposable entanglement 
witnesses. They are derived from the approach of PPT mixtures \cite{bastian}, 
and how this method is described in the following.

In this approach, instead of considering biseparable states, one
considers a superset of the set of biseparable states, the PPT 
mixtures. This set is given by those states that can be written 
as convex combination of states that have a positive partial 
transposition (PPT) with respect to a fixed bipartition \cite{peres96}. 
For three particles, these states are of the form
\begin{equation}
\varrho^{\rm{pmix}}
= p_{1}\varrho^{\rm{ppt}}_{A|BC}+p_{2}\varrho^{\rm{ppt}}_{C|AB}+p_{3}\varrho^{\rm{ppt}}_{B|AC}.
\end{equation}
Since every separable state is necessarily a PPT state, the PPT mixtures 
are indeed a superset of the biseparable states (see Fig.~1). In order to characterize
PPT mixtures in a better way, it was then shown in Ref.~\cite{bastian} that all 
states which are not PPT mixtures can be detected by entanglement 
witnesses with a simple form:

\noindent
{\bf Observation 1.}
Every state $\varrho$ that is not a PPT mixture can be detected
by a witness that can be written as
\begin{equation}
\WW = P_{M}+Q_{M}^{T_{M}} \mbox{ with } P_{M} \geq0,\;\; Q_{M}\geq 0,
\end{equation}
for all possible bipartitions $M$. Witnesses of this type are also positive
on all PPT mixtures. These witnesses are called fully decomposable witnesses. 

Therefore, the characterization of PPT mixtures boils down to a characterization
of fully decomposable witnesses. For a given multiparticle quantum state 
$\varrho$, the question whether it is detected by a fully decomposable 
witness can be formulated as the optimization problem \cite{bastian}
\begin{eqnarray}
\textrm{minimize:} &{\;\;}& Tr(\WW \varrho),
\nonumber
\\
\textrm{subject to:}
&{\;\;}&
Tr(\WW)=1~ \textrm{and for all partitions M:}
\nonumber
\\
&& \WW=P_{M}+Q^{T_{M}}_{M},~~ Q_{M}, P_{M}\geq 0.
\label{sdp}
 \end{eqnarray}
This problem is a convex optimization problem that can efficiently 
be solved by semidefinite programming and a ready-to-use implementation
which can in practice be applied to systems up to six qubits is freely available 
\cite{pptmixer}.

The approach of PPT mixtures for permutationally invariant states was 
recently investigated by Novo et al.~\cite{novo}.  With the help of the 
symmetry of such states, a simplified formulation was found, which leads 
to a numerical optimization problem, where the number of parameters scales 
only polynomially as $\mathcal{O}[N^{7}]$. This allows to investigate system 
sizes of about ten qubits for which the original implementation fails. Though 
this is an advance, numerical solutions are always restricted to a limited 
system size and analytical solutions are needed. This is exactly the aim 
of the present paper.

In the following, we mainly deal with fully PPT witnesses, which is a special 
kind of fully decomposable witnesses. Fully PPT witnesses have the additional
property that the operators $P_{M}$ equal zero for any bipartition. Thus, 
they are of the structure
\begin{equation} 
\WW=Q_{M}^{T_{M}} \mbox{ with }Q_{M}\geq0.
\label{pptwitness}
\end{equation}
A fully PPT witness is easier to characterize than a fully decomposable 
witness because it is sufficient to show that any partial transposition 
with respect to fixed bipartition is positive. Clearly, the class of fully
PPT witnesses does not detect as many states as the class of fully decomposable
witnesses. But, as we will see below, the resulting entanglement criteria
are often already better than all other constructions known so far.

\section{Projective entanglement witnesses for arbitrary Dicke states}

As mentioned above, projective entanglement witnesses for Dicke states 
are known analytically only for two special cases. In this section
we will fill this gap, but first we derive a useful result  on the 
structure of general projective entanglement witnesses.\footnote{We stress
that this statement has occurred in the literature, the implicit proof from
Ref.~\cite{graphbell} is, however, incomplete.}

\noindent
{\bf Proposition 2.}
Every projective entanglement witness is a fully PPT witness, therefore 
it detects only states which are NPT (that is, they are not PPT) 
with respect to all bipartitions.

{\it Proof.}
We consider the Schmidt decomposition of the state $|\psi\rangle$ with 
respect to a fixed bipartition
$
|\psi\rangle=\sum_{i}a_{i}|ii\rangle,
$
where the Schmidt coefficients are defined in decreasing order, 
i.e. $a_{0}\geq a_{1}\cdots\geq a_{N}.$
Then the corresponding projective entanglement witness is given by
$
\WW=\alpha \mathds{1}-|\psi\rangle\langle\psi|.
$
and the coefficient $\alpha$ fulfills $\alpha \geq a_0^2.$
Calculating the partial transposition of $\WW$ for the given bipartition
results in a matrix that can be brought, via permuting columns and 
rows, into block structure consisting of $2\times2$-blocks of the form
\[X=\left(
\begin{array}{cc}
\alpha &-a_i a_j\\
-a_i a_j &\alpha\\
\end{array}\right).
\]
These blocks have no negative eigenvalues since 
$\alpha \geq a_{i}a_j$ for all $i,j$ and therefore 
this partial transposition, as well as the transpositions for the 
other bipartitions, are non-negative.
\hfill$\Box$

Based on the Schmidt decomposition, we are able to calculate the projective 
entanglement witness for Dicke states with arbitrary $k$. We formulate it 
for any $k < N/2$, the 
expression in the case of $k={N}/{2}$ was already given in 
Ref.~\cite{gezajosa}, and the case
$k>N/2$ corresponds to the case $k' = N-k < N/2$.

\noindent
{\bf Proposition 3.}
For an arbitrary Dicke state $|D^{N}_{k}\rangle$ with $1< k<\frac{N}{2}$, 
the corresponding projective entanglement witness $\WW_{proj}$ is given by
\begin{equation}
\WW_{proj}=\frac{N-k}{N}\mathds{1}-|D^{N}_{k}\rangle\langle D^{N}_{k}|.
\end{equation}

{\it Proof.} We determine the Schmidt decomposition of $ |D^{N}_{k}\rangle$ 
for the bipartition $\hat A| \hat B$, where $\hat A$ denotes the first $A$ qubits, and $\hat B$ denotes the last $B=N-A$ qubits. We can rewrite the 
Dicke state as
\begin{eqnarray}
\nonumber
|D^{N}_{k}\rangle&=& 
\frac{1}{\sqrt{\binom{N}{k}}} 
\sum\limits_{j}P_{j}\big\{|1\rangle^{\otimes k}\otimes|0\rangle^{\otimes (N-k)}\big\}
\\
&=&
\frac{1}{\sqrt{\binom{N}{k}}} 
\sum\limits_{\alpha=0}^{k}\sqrt{\binom{A}{\alpha}}\sqrt{\binom{B}{k-\alpha}}
\;
|\alpha\rangle_{\hat A}\otimes |k-\alpha\rangle_{\hat B}.
\end{eqnarray}
Here, $|\alpha\rangle_{\hat A}$ (resp., $|k-\alpha\rangle_{\hat B}$)
denotes a normalized and symmetrized superposition of $\alpha$ (resp., 
$k-\alpha$) excitations on the system $\hat A$ (resp., $\hat B$); in other
words, $|\alpha\rangle_{\hat A}$ is the Dicke state $|D^{A}_{\alpha}\rangle$
on the first $A$ qubits. For the maximal overlap $\alpha$ with the biseparable
states, we now have to compute the maximally possible squared Schmidt coefficient,
\begin{equation}
\alpha =\max_{1 \leq A\leq N-1} \;\;\; \max_{0\leq \beta \leq k}
\frac{\binom{A}{\beta} \binom{N-A}{k-\beta}}{\binom{N}{k}}
=
\frac{N-k}{N}.
\end{equation}
Details of this maximization are described in Lemma 10 in Appendix A1.
\hfill$\Box$

We note that the projective witness can be improved since the full
identity operator $\mathds{1}$ is not needed in order to guarantee positivity on
all biseparable states. Indeed, as already known for the W state 
\cite{hartmut}, one can substitute the $\mathds{1}$ by $\mathds{1}_{2k}$ which 
is the identity on the space with maximally $2k$ excitations. Using the 
same argument as in the proof of Proposition 2 one finds that also with this substitution
the witness is a fully PPT witness. So the refined projective witness is 
given by
\begin{equation}
\WW'_{proj}=\frac{N-k}{N}\mathds{1}_{2k}-|D^{N}_{k}\rangle\langle D^{N}_{k}|.
\label{projbesser}
\end{equation}

Entanglement witnesses are frequently used tool in experiments, 
and in order to compare the strength of different witnesses, the
resistance to white noise is often used. For that, one considers
noisy states of the form 
\begin{equation}
\varrho_{\rm noise} = p\frac{\mathds{1}}{2^{N}}+(1-p)\ketbra{D^{N}_{k}},
\end{equation}
and asks for the critical value $p_{\rm crit}$ when the state 
is not detected anymore. 
For the witness $\WW'_{proj}$ the noise tolerance is 
\begin{equation}  
p_{\rm crit} = \frac{1}{1+\frac{1}{2^{N}}
\big[\frac{N-k}{k}\sum\limits_{i=0}^{2k}\binom{N}{i}-\frac{N}{k}\big]}.
\label{projnoise}
\end{equation}
For any $k <N/2$ the value of $p_{\rm crit}$ tends to one if particle 
number $N$ increases, meaning that a large fraction of white noise
can be added and still the entanglement is preserved. Therefore, 
the genuine multiparticle entanglement in Dicke states with $k <N/2$ 
is very robust for large particle numbers. 

\section{Advanced entanglement witnesses for general 
Dicke states}
In this section, we will present our main idea to construct strong entanglement
criteria for Dicke states. To explain our idea, we start with a discussion of
the Dicke state with $k={N}/{2}$ excitations and derive a simple improvement
of the projector based witness in Eq.~(\ref{gezaproj}). Then, we introduce a general
construction for arbitrary Dicke states. 

To start, we make the following ansatz for a witness for Dicke states 
with $k={N}/{2}$ excitations: 
\begin{equation}
\label{n2ansatz}
\WW_{\frac{N}{2}}
=
\sum_{i=0}^{N}\omega_{i}\Pi_{i}-|D^{N}_{\frac{N}{2}}\rangle\langle D^{N}_{\frac{N}{2}}|.
\end{equation}
Here, $\Pi_{i}$ is the projector onto the space with exactly $i$ excitations. The task
is now to determine the parameters $\omega_i$ such that the witness is positive on all
biseparable states. One possible choice is to take $\omega_i = N/(2N-2)$, since this
leads to the witness in Eq.~(\ref{gezaproj}). For some values of $i$ one can choose 
smaller $\omega_i$, however, and improve the witness in Eq.~(\ref{gezaproj}) in this 
way. 

To do so, the parameters $\omega_{i}$ are determined by the requirement 
that all partial transpositions of $\WW_{\frac{N}{2}}$ should be positive
semidefinite [see Eq.~(\ref{pptwitness})].  Since the calculation of the 
partial transposition of the witness turns out to be demanding, we first 
use an approximation known as Gershgorin circles \cite{gershgorin}.

\noindent
{\bf Lemma 4.}
Let $A=(a_{ij})$ a complex $n\times n$-matrix, 
$R_{i}=\sum_{j \neq i}|a_{ij}|$ the sum of the off-diagonal elements
in the $i$-th row, and $D_i = D(a_{ii},R_{i})$ the closed disc  around 
$a_{ii}$ with radius $R_i$. Then, all eigenvalues of $A$ lie in the
union of the discs $D_i$.

Since the eigenvalues of a Hermitean matrix $A$ are real, it is sufficient 
that all Gershgorin circles lie in the half space with positive real 
part in the complex plane  in order  to guarantee that $A$ is positive 
semidefinite.
This will be used in the following.

\noindent
{\bf Proposition 5.}
The operator  $\WW_{\frac{N}{2}}$ is a fully PPT entanglement 
witness if
\begin{equation}
\omega_{i}=\min
\left\{
\frac{\binom{N}{i}}{\binom{N}{{N}/{2}}},\frac{1}{2}\frac{N}{N-1}
\right\}.
\label{betterchoice}
\end{equation}
holds for all $i=0,1,...,N$.

{\it Proof.} The condition follows by calculating the partial transposition of 
$\WW_{\frac{N}{2}}$ with respect to the first $\delta$ qubits, where $\delta$
is fixed but arbitrary in the interval $1\leq\delta\leq \frac{N}{2}$. This 
transposition has a block structure where the blocks 
are of the form
\be
X_\delta =
\left(\begin{array}{cc}
  \omega_{N/2+\delta}\mathds{1}&A\\
   A& \omega_{N/2-\delta}\mathds{1}\\
  \end{array}\right).
  \label{testgleichung}
\ee
Here, all the four sub-blocks  of the matrix $X_\delta$ are 
$\binom{N}{i}\times\binom{N}{i}$ matrices (with $i=N/2+\delta$ 
or $i=N/2-\delta$) and the entries in the block $A$ are zero or $1/{N \choose N/2},$
coming from the normalization of the state $|D^{N}_{\frac{N}{2}}\rangle.$ Then, Lemma
4 implies that if $\omega_{i}\binom{N}{{N}/{2}} \geq \binom{N}{i}$
 is fulfilled, then the matrix $X_\delta$ 
is positive semidefinite. 
The choice in Eq.~(\ref{betterchoice}) is then justified because we know from 
Proposition~1 that the 
projective witness in Eq.~(\ref{gezaproj}) is also a fully PPT
witness. Therefore, also for the choice $\omega_i = N/(2N-2)$ the 
block matrices $X_\delta$ are positive semidefinite. Clearly, taking
the minimum of both expressions results in the strongest witness. 
\hfill $\Box$

To generalize the ansatz of Eq.~(\ref{n2ansatz}) to arbitrary Dicke states, we make the ansatz
\begin{equation}
\WW^{opti}_{N,k}=\sum\limits_{i=0}^{2k}\omega_{i}\Pi_{i}-|D^{N}_{k}\rangle\langle D^{N}_{k}|
\label{dickeansatz}
\end{equation}
for a fully PPT entanglement witness for Dicke states with $k\neq \frac{N}{2}$. To 
obtain a fully PPT witness, the parameters $\omega_{i}$ have to fulfill certain 
conditions which can be found by a detailed singular value decomposition of 
matrices similiar to
matrix $A$ in Eq.~(\ref{testgleichung}). One finds:

\noindent
{\bf Theorem 6.} The operator $\WW^{opti}_{N,k}$ is a fully PPT witness, if
the following two conditions are fulfilled:

\noindent
(i) for $ i \neq k$ the parameters $\omega_{i}$ obey
\be
\omega_{k-\delta}{\omega_{k+\delta}} \geq (\lambda_{\rm max})^2 := \max_{x_1, x_2}
\{\lambda^2(x_1,x_2)\},
\label{thm6a}
\ee
where the coefficients $\lambda(x_1,x_2)$ are given by
\bea
\lambda^2(x_1,x_2) &=& 
{N-x_1 -x_2 \choose k-x_2}
{x_1+x_2 \choose x_2}
{N- x_1 - x_2 \choose k-\delta - x_2}
{x_1+x_2 \choose x_2+\delta} 
/{N \choose k}^2.
\nonumber
\\
\eea
For the maximization of $\lambda^2(x_1,x_2)$, it suffices to consider 
the case that $x_1 \geq \delta$ and $x_1 +x_2 \leq N/2.$

\noindent
(ii) for $ i = k$ the parameter $\omega_{i}$ is given by 
\be
\omega_{i} = \mu (N,k)
:=
\left\{
\begin{array}{l}
\frac{N-k}{N} \mbox{ for $k<N/2$}
\\
\frac{N}{2(N-1)} \mbox{ for $k=N/2$}
\end{array}
\right. .
\ee

{\it Proof.} The proof for this statement can be found in Appendix A2.
It should be noted that the alternative estimate 
$\omega_{k\pm\delta} \geq \mu(N,k)$, as it was used 
in Eq.~(\ref{betterchoice}), is not useful here, and 
the estimate from Eq.~(\ref{thm6a}) is always stronger. 
This follows from the previous result in Proposition 3, 
see also Eqs.~(\ref{nice1}, \ref{nice2}) in the Appendix A2.
Applied to the definition of $\lambda^2(x_1, x_2)$ these
estimates show that $\lambda^2(x_1, x_2) \leq \mu^2(N,k).$
\hfill $\Box$

\begin{figure}[t]
\centering
\includegraphics[width=0.66\textwidth]{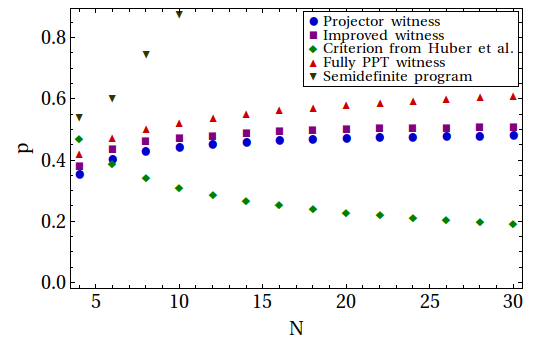}
\caption{Comparison of the noise robustness of the different entanglement
detection methods for Dicke states with $k={N}/{2}$. See text for further
details.}
\end{figure}

It remains to find the optimal choice of the coefficients $\omega_i$
in order to maximize the noise tolerance. If we consider white noise, 
the optimal noise tolerance is  achieved, if the trace of $\WW^{opti}_{N,k}$ 
is minimal. This is the case if for any $\delta \geq 1$ the value
${N \choose k-\delta}\omega_{k-\delta} + {N \choose k+\delta}\omega_{k+\delta}$
is minimal. Using Lagrange multipliers one can directly derive the following
result.

\noindent
{\bf Corollary 7.} Consider the witness $\WW^{opti}_{N,k}$  
from Theorem 6. The optimal noise robustness  with respect to white noise
is achieved if the $\omega_{k\pm\delta}$ are given by
\be
\omega_{k-\delta} =\sqrt{\frac{{N \choose k+\delta}}{{N \choose k-\delta}}}
\lambda_{\rm max}
\;\;\;
\mbox{  and  }
\;\;\;
\omega_{k+\delta} =\sqrt{\frac{{N \choose k-\delta}}{{N \choose k+\delta}}}
\lambda_{\rm max}.
\ee

Now it is time to compare our new criteria with existing criteria from 
the literature. Here, the work of Huber et al.~\cite{huber} is important,
where interesting entanglement criteria for Dicke states have been derived, 
which are for many cases the best known results thus far. The noise tolerance 
of these criteria is given by
\begin{equation}
p_{\rm crit} =
\frac{1}{1+
\frac{1}{2^N}
\big[
(2N-2k-1)\binom{N}{k}
\big]
}.
\end{equation}
In addition to these criteria we consider for $N\leq 10$ also the direct solution
of  the semidefinite program in Eq.~(\ref{sdp}), obtained with the methods of 
Ref.~\cite{novo}.

In Fig.~2 we compare the criteria for the case $k=N/2$. It can be seen that the 
improved witness from Proposition~5 
is better than the projective witness by some amount, but the fully PPT 
witness from Theorem 6  results in a 
further significant improvement. The full solution of the semidefinite 
program is, for the cases that it can be computed, clearly the best 
criterion. This shows that there is still space for improving the 
analytical criteria of this paper. 

\begin{figure}[t]
\centering
\subfigure[]{\includegraphics[width=0.45\textwidth]{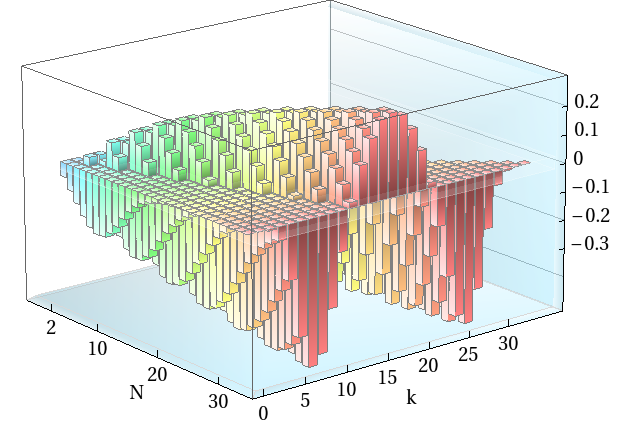}} 
\hspace*{0.5cm}
\subfigure[]{\includegraphics[width=0.40\textwidth]{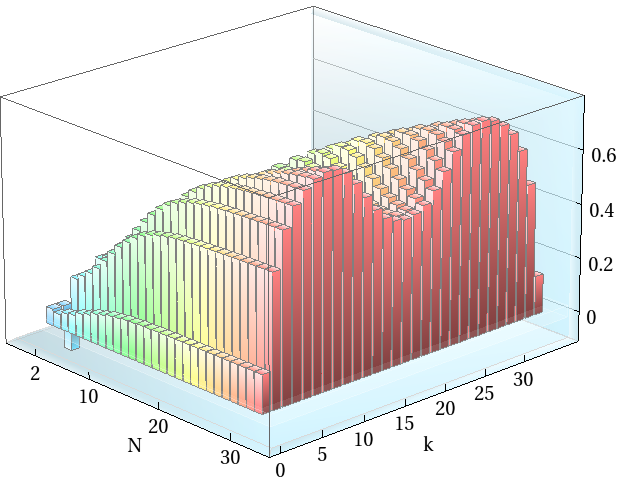}}
\caption{(a) Comparison between the projective entanglement witness 
[Proposition 3 and Eqs.~(\ref{projbesser}, \ref{projnoise})] and the 
criterion of Ref.~\cite{huber}. Here, the difference of the noise robustnesses 
$p_{\rm crit}$ is shown. For the cases with $k\approx N/2$, Eq.~(\ref{projbesser})
results in a better criterion, if $k \ll N/2$  or $k \gg N/2$ the results from 
Ref.~\cite{huber} are stronger.
(b) Comparison between the PPT entanglement witness (Theorem 6) 
and the criterion of Ref.~\cite{huber}. Here, for all cases apart from 
$(N,k)=(4,2)$ the criteria from Theorem 6 are stronger. 
The figure displays $(p_{\rm crit}^{\rm Thm. 6} - p_{\rm crit}^{\rm Ref. [9]})/
(1-p_{\rm crit}^{\rm Ref. [9]})$. This quantifies the fraction of the states 
not detected by Ref~\cite{huber}, which are in addition detected by Theorem 6. 
}
\end{figure}

In Fig.~3(a) we compare the  projective witness from Proposition~3 
and Eqs.~(\ref{projbesser}, \ref{projnoise}) with the results of 
Ref.~\cite{huber}. For the case $k\approx N/2$, the criterion from 
Eq.~(\ref{projbesser}) is better, but if $k \ll N/2$  or $k \gg N/2$ 
the results from Ref.~\cite{huber} are superior. In Fig.~3(b) we compare
the PPT entanglement witness (Theorem 6) and the criterion of 
Ref.~\cite{huber}. Here, for all cases [apart from $(N,k)=(4,2)$] the 
criteria from Theorem 6 are stronger. 
Finally, a detailed comparison
for the Dicke states with $k=1$ can be found in the following section.

\section{Results for W states}

In this section, we discuss in some detail entanglement witnesses 
for Dicke states with only one excitation, the so-called W states. 
Besides the construction from Theorem 6 we will introduce a different
type of witnesses, which do not rely on the approach of PPT mixtures. 

The set of W states is given by
$
\ket{W_N}
=
|D^{N}_{1}\rangle
=
(|100\dots0\rangle+|010\dots0\rangle+\dots+|000\dots1\rangle)/{\sqrt{N}}.
$
For these states, we use the same ansatz for a fully PPT entanglement 
witness  as in the section before (see Eq.~(\ref{dickeansatz})),
\begin{equation}
\WW^{opti}_{N}=\sum\limits_{i=0}^{2}\omega_{i}\Pi_{i}-\ketbra{W_N}.
\label{wansatz}
\end{equation}
Note that it was shown in Ref.~\cite{localmeas} that for the 
experimental determination of $\WW^{opti}_{N}$ only $2N-1$ 
local measurements are necessary. First, we can now explicitly 
write down the witnesses following the method in Theorem 6. 
We have:
\newpage
\noindent
{\bf Corollary 8.} Consider the $N$-qubit W state. If $N$ is even,
the witness with the highest noise tolerance according to Corollary 7
is given by 
\begin{equation}
\omega_{0}=\frac{\sqrt{N(N-1)}}{\sqrt{8}},
\;\;\;
\omega_{1}=\frac{N-1}{N},
\;\;\;
\omega_{2}=\frac{1}{\sqrt{2 N (N-1)}}.
\end{equation}
For the case that $N$ is odd, the optimal coefficients are
\begin{equation}
\omega_{0}=\frac{\sqrt{(N-1)^2(N+1)}}{\sqrt{8N}},
\;\;\;
\omega_{1}=\frac{N-1}{N},
\;\;\;
\omega_{2}=\frac{\sqrt{N+1}}{\sqrt{2 N^3}}.
\end{equation}
For both cases, the noise tolerance is given by
\be
p_{\rm crit} = 
\frac{1}{1+\frac{1}{2^N}\big[2 N \omega_0 +N(N-2)\big]}.
\ee

{\it Proof.} Starting from Theorem~6, we have to compute 
$\lambda_{\rm max}^2.$ In the definition of $\lambda^2(x_1, x_2)$, 
we have $k= \delta =1$ and therefore $x_2=0.$ Then, 
$\lambda^2(x_1, 0) = x_1(N-x_1)/N^2.$ If $N$ is even, 
this is maximal for $x_1=N/2$ and therefore 
$\lambda_{\rm max}^2 = 1/4.$ If $N$ is odd, the optimum is
attained at $x_1=(N-1)/2,$ resulting in $\lambda_{\rm max}^2 = 1/4 (1-1/N^2).$
Then the claim follows from a simple calculation. 
\hfill $\Box$

We can improve this result by going back to the original definition 
of an entanglement witness as an observable being positive on all biseparable 
states. Using the ansatz in Eq.~(\ref{wansatz}), the parameters $\omega_{i}$ 
can be determined by requiring that the minimal overlap of $\WW_{N}^{opti}$ 
with the pure biseparable states is non-negative. This defines a new 
witness $\widetilde{\WW}_{N}$. In the following, we show that this 
optimization problem can be reduced to a simple numerical optimization, 
in principle, also other methods might be feasible \cite{sperling}.
Note that the following proposition is a generalization of the method
used in Ref.~\cite{hartmut} to construct witnesses for W states. 

\noindent
{\bf Proposition 9.}
For given parameters $\omega_{0},\omega_{1}$ and 
$\omega_{2}$, the operator $\widetilde{\WW}_{N}$ defines an entanglement witness iff the
result of the following optimization, 
\begin{eqnarray}
\Upsilon &:=&
\max_{K,L} 
\max_{\alpha,\beta} 
\max_{\delta,\gamma} 
\Big\{
\frac{(K\beta\gamma+L\delta\alpha)^{2}}{N}-
\omega_{0}\alpha^{2}\gamma^{2}-
\nonumber
\\
&&
\omega_{1}(\alpha^{2}+\gamma^{2}-2\alpha^{2}\gamma^{2})
-\omega_{2}(1-\alpha^{2}-\gamma^{2}+\alpha^{2}\gamma^{2})
\Big\}
\end{eqnarray}
is not positive, $\Upsilon \leq 0.$ Here, the optimization 
must be carried out under the constraints
\begin{equation}
K+L=N,
\;\;\;
\alpha^{2}+K\beta^{2}=1
\;\;\;
\mbox{ and }
\;\;\;
\gamma^{2}+L\delta^{2}=1.
\end{equation}
{\it Proof.} The proof is given in the Appendix A3.
\hfill $\Box$

Before comparing this result with the previous witnesses, 
let us discuss its implications. First, note that the 
maximizations in the definition of $\Upsilon$ are lengthy, 
but elementary. They may be rewritten as a linear optimization 
with polynomial constraints, then they belong to a class of 
problems, where the {\it global} maximum can be obtained in 
an efficient manner \cite{lasserre}. Also, the complexity of the 
optimization procedure does not increase 
with the number of qubits.

Second, note that shifting all the $\omega_i$ by a 
constant amount, i.e. 
$\omega_i \rightarrow \omega_i+\varepsilon$
results in a decrease of $\Upsilon$ by the same 
constant $\varepsilon.$ Hence, Proposition 9
constitutes a constructive way to determine the 
$\omega_i$ for an good witness: One starts with arbitrary 
$\omega_i,$ performs the maximization resulting 
in a constant $\tilde \Upsilon,$ then shifts the 
$\omega_i$ by  $\tilde \Upsilon$ to achieve 
that finally $\Upsilon=0.$

\begin{figure}[t]
\centering
\includegraphics[width=0.66\textwidth]{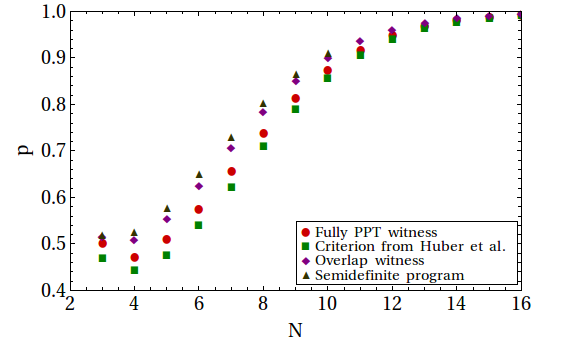}
\caption{Comparison of the different criteria (Corollary 
8, Proposition 9 and Refs.~\cite{huber, novo}) for $N$-qubit W states. 
See the text for further details.}
\end{figure}

In Fig.~4 we compare the witness from Corollary 8, the criterion from
Ref.~\cite{huber}, the overlap witness from Proposition 9 and the solution
of the semidefinite program \cite{novo}. One can clearly 
see that Proposition 9 results in a stronger criterion than Corollary 8.
This is not surprising: The entanglement criterion from Proposition  is 
more general than the fully PPT entanglement witness, since it does not 
rely on any structural approximation. It should be noted, however, that 
for large $N$ all three criteria detect nearly all W states affected by
white noise. 

\section{Conclusion}
Using the approach of PPT mixtures we have derived entanglement criteria
which are suited for Dicke states. In most cases, these criteria are stronger
than entanglement conditions previously known. In this way, we have demonstrated
how the method of PPT mixtures, being at first sight a numerical tool, can be 
used to gain analytical insight into the structure of entanglement.

There are several ways to extend our results. First, one can try to 
derive analytical conditions for an operator as in Eq.~(\ref{dickeansatz})
to be a fully decomposable witness. This, however, seems very demanding to 
us. A more direct research problem is the application of the method of PPT 
mixtures to other families of states with symmetries, e.g. states with 
$U \otimes U \otimes ... \otimes U$-symmetry \cite{uuusymmetry}. This may finally help
to answer the question for which types of states the criterion of PPT mixtures
constitutes a necessary and sufficient criterion of genuine multiparticle entanglement. 

We thank Matthias Christandl, Tobias Moroder and Leonardo Novo for 
discussions and an anonymous referee for very helpful remarks.
This work has been supported by the EU (Marie Curie CIG 293993/ENFOQI) 
and the BMBF (Chist-Era Project QUASAR).

\section*{Appendix}

\subsection*{A1: A Lemma required for the proof of Proposition 3}
\noindent
{\bf Lemma 10.}
Consider a pair of integer numbers $A,B \geq 1$ with $A+B=N$ 
and a second pair $\alpha,\beta \geq 0$ with $\alpha+\beta=k$, 
where $k< N/2.$ Then the inequality
\begin{equation}
\binom{A}{\alpha}\binom{B}{\beta} \leq \binom{N-1}{k}
\label{lemmax1}
\end{equation}
holds, and equality is assumed for $A=1, B=N-1$
and $\alpha=0$ and $\beta=k.$

{\it Proof.}\footnote[8]{From an anonymous referee, we learned that 
our proof of Lemma 10 can be made significantly shorter. First, one can 
assume without loss of generality that $\alpha < A/2$ and consider 
the case $\alpha \geq 1, \beta \geq 1$ Then, one has the estimate
$
\binom{A}{\alpha}\binom{B}{\beta}
=
\binom{A}{\alpha}\binom{B-1}{\beta}
+
\binom{A}{\alpha}\binom{B-1}{\beta-1}
\leq 
\binom{A}{\alpha}\binom{B-1}{\beta}
+
\binom{A}{\alpha+1}\binom{B-1}{\beta-1}
<
\binom{N-1}{k},
$
where also Vandermonde's identity \cite{chinese} has been used. 
This proves already the claim.
}
First, consider the case that $A=1$ (or, equivalently, $B=1$). 
Then, for $\alpha$ only the possibilities $\alpha=0$ or $\alpha=1$
have to be considered, but for both cases the statement in Eq.~(\ref{lemmax1})
is clear. Therefore, we can assume in the following that $A\geq 2$ and 
$B \geq 2.$
Then, using the fact that
\be
\binom{N-1}{k}
=\binom{N}{k}-\binom{N-1}{k-1}
=\binom{N}{k}-\left[\binom{N-2}{k-2}+\binom{N-2}{k-1}\right]
\ee
the statement of the Lemma is equivalent to
\begin{equation}
\underbrace{\binom{A}{\alpha}\binom{B}{\beta}}_{T_1}
+
\underbrace{\binom{N-2}{k-1}}_{T_2}+
\underbrace{\binom{N-2}{k-2}}_{T_3}
\leq\binom{N}{k}.
\label{lemmax2}
\end{equation}
Since $k < N/2$ we have that
\begin{equation}
\frac{\binom{N-2}{k-2}}{\binom{N-2}{k-1}}=\frac{k-1}{N-k} < 1,
\end{equation}
which implies that $T_2 > T_3$ and consequently $T_3 \leq T_2 - 1.$

In the following, we will prove Eq.~(\ref{lemmax2}) using a combinatorial
argument. The right hand side of Eq.~(\ref{lemmax2}) is the number of 
possibilities to choose $k$ elements out of a set $\hat{N}$ with $N$ elements, and we show that 
each of the three terms appearing on the left hand side of Eq.~(\ref{lemmax2}) can be 
interpreted as the number of {\it different} ways to choose $k$ objects out 
of $N$. To start,  we consider a partition of the set $\hat N$, 
i.e.  two disjoint subsets $\hat A$ and $\hat B$ satisfying 
$\hat A \cup \hat B = \hat N$ with the cardinalities
$|\hat A| = A$ and $|\hat B| = B.$ We proceed in three steps.

{\it Step 1.} The first term  $T_1$ equals the number of ways to choose $\alpha$ elements 
out of $\hat{A}$ and $\beta$ elements out of $\hat{B}$. 

{\it Step 2.} To catch the second term
$T_2$, we fix two elements, $a_{0} \in \hat A$ and $b_{0}\in \hat B$ and choose 
$k-1$ objects out of the remaining $N-2$ objects. This results in $a$ elements 
out of $\hat A$ and $b$ elements out of $\hat B$ with $a+b=k-1$. 
Due to the fact that $a+1=\alpha$ and $b+1=\beta$ cannot hold at the same time, 
we can choose to add either $a_{0}$ or $b_{0}$ (leading to a selection of $k$ 
elements out of $N-1$), which is different from the selections in Step 1, since 
now a different number of terms from $\hat A$ and $\hat B$ is chosen.
The number of different selections obtained in that way is equal to the 
second term $T_2$.

{\it Step 3.} In order to justify the third term, a more detailed analysis is required. Let us
start with a given choice of a subset $\hat a$ with $a$ elements out of 
$\hat A \setminus \{a_0\}$ and a subset $\hat b$  with $b$ elements out 
of $\hat B \setminus \{b_0\}$, which we used in Step~2. Note that 
since $A \geq 2$ and $B\geq 2$ the sets
$\hat A \setminus \{a_0\}$ and $\hat B \setminus \{b_0\}$ are not 
empty. We can distinguish two cases:
\begin{enumerate}

\item If $a \geq 1$ and $b\geq 1$, then the sets $\hat a$ and $\hat b$ are not empty, 
and we can choose elements $a_+ \in \hat a$ and $b_+ \in \hat b$. Then, we can
interchange both $a_+$ with $a_0$ and $b_+$ with $b_0$ and afterwards we can decide
(as in Step 2) to add again $a_+$ or $b_+$ to arrive at a selection of $k$ elements
out of $N$. This choice will be different from the one in Step~2, since now both
$a_0$ and $b_0$ are in. It will also be different from the  choice in Step~1, since
we can choose $a_+$ or $b_+$, similar as in Step 2.

\item If $a \neq A-1$ and $b \neq B-1$, then the sets $\hat a$ and $\hat b$ 
are not equal to the respective supersets $\hat A \setminus \{a_0\}$ and 
$\hat B \setminus \{b_0\}.$ Then we can choose elements 
$a_- \in \hat A \setminus (\{a_0\}\cup\hat a)$ 
and 
$b_- \in \hat B \setminus (\{b_0\}\cup\hat b)$ 
and interchange both $a_-$ with $a_0$ and $b_-$ with $b_0.$
Afterwards, we can decide (as in Step 2) to add $a_-$ or $b_-$ 
to arrive at a selection of $k$ elements out of $N$. Again, this 
choice will be different from the one in Step~2, since now neither
$a_0$ nor $b_0$ are in. It is also different from the choice in Step 1.
\end{enumerate}
For a given choice of $\hat a$ and $\hat b$ these cases are not exclusive, 
in fact, often both cases are true. The only situation where {\it none} of
the cases applies is if $a = A-1$ and simultaneously $b = 0$ (or vice versa),
corresponding to the case that when choosing $k-1$ elements out of 
$\hat{N}\setminus\{a_0, b_0\}$ in Step 2, the complete set $A \setminus \{a_0\}$
was chosen. This can only happen if $A=k.$ Note, however, that then
only for a single example of the different choices in Step 2 none of 
the conditions in Step 3 applies, for all other choices at least one 
of the cases is true.

This leads to our final estimate: If $A \neq k$ and $B \neq k$
we can apply all the three steps outlined above, leading to 
$T_1 + T_2 +T_2 \leq {N \choose k}$, which implies Eq.~(\ref{lemmax2}).
If $A = k$ or $B = k$ the same procedure results in
$T_1 + T_2 +(T_2-1) \leq {N \choose k},$ also proving Eq.~(\ref{lemmax2}).
\hfill $\Box$

As a final remark, note that from our discussion it follows that
in the situation of Lemma 10 for $k<N/2$ the estimate
\be
\binom{A}{\alpha}\binom{B}{\beta}/\binom{N}{k} \leq \frac{N-k}{N}
\label{nice1}
\ee
follows, while for $k=N/2$ from Ref.~\cite{gezajosa} the bound
\be
\binom{A}{\alpha}\binom{B}{\beta}/\binom{N}{k} \leq \frac{N}{2(N-1)}
\label{nice2}
\ee
is known. These properties are useful for the construction of general 
PPT witnesses for Dicke states, see also the discussion in the proof 
of Theorem 6.

\subsection*{A2: Proof of Theorem 6}

As in the case of Proposition 5, the partial transposition of the 
witness $\WW^{opti}_{N,k}$ can, after appropriate reordering, written
in a block diagonal form, and the blocks are of the form
\be
X_\delta^{(k)} =
\left(\begin{array}{c c}
  \omega_{k-\delta}\mathds{1}&A\\
   A^T& \omega_{k+\delta}\mathds{1}\\
  \end{array}\right).
\ee
We have to prove that this matrix is positive semidefinite and 
for that we first characterize the off-diagonal block $A$. Note 
that here the matrix $A$ is rectangular and not quadratic as in the 
proof of Proposition 5. In the end, we want to determine the 
singular value decomposition of $A$, but this requires several 
steps.

{\it Step 1.} Let us start with the first row of $A$. The entries in 
this first column are of the type
\be
\ket{\chi}\bra{\eta}
=
\ket{0 \cdots 0 
\underbrace{
\underbrace{0 \cdots 0}_{x_1}
\underbrace{1 \cdots 1}_{x_2}}_{x}
\underbrace{1 \cdots 1}_{k-\delta - x_2}
}
\bra{\eta}.
\ee
This notation should be understood as follows: the vector $\ket{\chi}$
contains $k-\delta$ entries ``1'', which are, since we are considering
the  first row of $A$, aligned on the left. The partial transposition 
affects the qubits denoted by $x$. In the first $x_1$ of these qubits, 
the vector $\ket{\chi}$ has the entries ``0'', while for the last $x_2$
qubits of $x$ the vector $\ket{\chi}$ has the entries ``1''. In the following, 
it will be useful to denote a set of qubits by $\hat x$, and the corresponding
number of qubits by $x = |\hat x|.$
Clearly, we are considering a special type of the transposition, since the 
set $\hat x$ consists of neighboring qubits, but, as  we will see, this is no
restriction, and in the end only the numbers $x_1$ and $x_2$ matter, 
not their position.

Now we consider the possible values for $\bra{\eta}.$ Clearly, 
$\bra{\eta}$ has $k+\delta$ entries with the value ``1''. These
entries must be distributed in such a way, that if the transposition
on the qubits $\hat x$ is applied a second time on $\ket{\chi}\bra{\eta}$,
(that is, one considers $\ket{\chi'}\bra{\eta'}=(\ket{\chi}\bra{\eta})^{T_x}$)
then $\ket{\chi'}$ and $\bra{\eta'}$ have both $k$ entries with the 
value ``1''. Let $\alpha$ be the number of ``1'' in the set $\hat{x}_1$ for 
$\bra{\eta}$ and let $\beta'$ be the number of ``0'' in the set $\hat x_2$ 
for $\bra{\eta}.$ The condition on $\ket{\chi'}\bra{\eta'}$ is fulfilled, 
iff $\alpha- \beta' = \delta.$ There are ${x_1 \choose \alpha}{x_2 \choose \beta'}$ possibilities for that, and ${N-x_1 -x_2 \choose k+\delta-\alpha-(x_2-\beta')}$ possibilities for distributing the remaining
$[k+\delta-\alpha-(x_2-\beta')]$ entries ``1'' on the remaining $N-x_1 -x_2$
qubits. Since $\alpha$ is not fixed, the total number of possible
$\bra{\eta}$ is given by 
\begin{eqnarray}
n (x_1, x_2) &=& \sum_{\alpha}^{} 
{N-x_1 -x_2 \choose k+\delta-\alpha-(x_2-\beta')}
{x_1 \choose \alpha}{x_2 \choose \beta'}
\nonumber \\
&=&{N-x_1 -x_2 \choose k-x_2}
\sum_{\alpha}^{} {x_1 \choose \alpha} {x_2 \choose x_2 - \beta'}
\nonumber \\
&=& {N-x_1 -x_2 \choose k-x_2}
\sum_{\alpha}^{} {x_1 \choose \alpha} {x_2 \choose x_2 +\delta - \alpha}
\nonumber \\
&=& {N-x_1 -x_2 \choose k-x_2}
{x_1+x_2 \choose x_2+\delta},
\end{eqnarray}
where we have used Vandermonde's identity \cite{chinese} in the last step.
The number $n(x_1, x_2)$ is the number of entries in the first row of 
$A$, for the given partition.

Summarizing the first step, we can write the first row of $A$ as 
$\ket{\phi}\bra{\psi}$, where $\bra{\psi}$ is the sum over all possible
$\bra{\eta}.$ The norm of $\bra{\psi}$  is 
$\Vert \bra{\psi}\Vert = \sqrt{n(x_1, x_2)}/{N \choose k}$, since  
$1/{N \choose k}$ is the global factor in the entries of $A$, originating from the normalization of the Dicke state. 

{\it Step 2.} In the previous step we characterized the possible 
$\bra{\eta}$ (or their sum $\bra{\psi}$) for the first row. However, 
many other rows lead to the same $\bra{\psi}.$ Indeed, let us consider
a fixed set $\hat x$, where the partial transposition is applied to. 
A given row of $A$ is labeled by a vector $\ket{\chi}$ with $k-\delta$
entries equal to ``1''. We can define the set $\hat x_1$ as the subset of
$\hat x,$ where the entries of $\ket{\chi}$ are ``0'', and $\hat x_2$ 
as the subset of $\hat x,$ where the entries of $\ket{\chi}$ are ``1''.
All rows which lead to the same sets $\hat x_1$ and $\hat x_2$ lead to
the same $\bra{\psi}$ in Step 1. For given $\hat x_1$ and $\hat x_2$
there are ${N- x_1 - x_2 \choose k-\delta - x_2}$ rows compatible
with that. Therefore, we can sum over these rows, and can write $A$ as
\be
A = \sum_{\hat x_1 \cup \hat x_2 = \hat x} 
\mu(\hat x_1,\hat x_2)
\ket{\phi(\hat x_1, \hat x_2)}\bra{\psi(\hat x_1 ,\hat x_2)}.
\ee
Here, the vectors $\ket{\phi(\hat x_1, \hat x_2)}$ and 
$\bra{\psi(\hat x_1 ,\hat x_2)}$ are normalized, and the coefficients
are given by 
$
\mu(\hat x_1,\hat x_2) = 
\sqrt{n(x_1, x_2) {N- x_1 - x_2 \choose k-\delta - x_2}}/{N \choose k}.
$
The vectors $\ket{\phi(\hat x_1, \hat x_2)}$ are clearly orthogonal, 
the vectors $\bra{\psi(\hat x_1 ,\hat x_2)}$, however, are not yet orthogonal.

{\it Step 3.} Now we describe the orthogonality relations of the vectors
$\bra{\psi(\hat x_1 ,\hat x_2)}$ in some more detail. First, we consider
two vectors $\bra{\psi(\hat x_1 ,\hat x_2)}$ and 
$\bra{\psi(\hat{\underline{x}}_1, \hat{\underline{x}}_2)}$ and prove 
that  they are orthogonal, if $x_1 \neq \underline{x}_1$ (and 
consequently $x_2 \neq \underline{x}_2$). Consider the terms $\bra{\eta}$
as discussed in Step 1 for $\bra{\psi(\hat x_1 ,\hat x_2)}$ and 
$\bra{\psi(\hat{\underline{x}}_1, \hat{\underline{x}}_2)}$ and assume that
two of them ($\bra{\eta}$ from $\bra{\psi(\hat x_1 ,\hat x_2)}$ and 
$\bra{\underline\eta}$ from $\bra{\psi(\hat{\underline{x}}_1, \hat{\underline{x}}_2)}$) 
are {\it not} orthogonal. Let 
$\alpha$ the number of ``1'' in $\hat x_1$ in $\bra{\eta}$, 
$\alpha'$ the number of ``0'' in $\hat x_1$ in $\bra{\eta}$, 
$\beta$  the number of ``1'' in $\hat x_2$ in $\bra{\eta}$, 
$\beta'$ the number of ``0'' in $\hat x_2$ in $\bra{\eta}$,
and let $\underline\alpha$, $\underline\alpha'$,
$\underline\beta$,
$\underline\beta'$ be the analogous quantities for 
$\bra{\underline\eta}$.
If $\braket{\eta}{\underline\eta} \neq 0$ we must have 
$\alpha + \beta = \underline \alpha + \underline \beta$
and 
$\alpha' + \beta' = \underline \alpha' + \underline \beta'$,
together with $\alpha-\beta' = \delta = \underline \alpha - \underline \beta'$
this implies that $\beta - \alpha'= \underline \beta - \underline \alpha'.$
On the other hand, let us assume that 
$\underline{x}_1 =  \underline \alpha + \underline \alpha' > x_1 = \alpha + \alpha'.$ Adding $\beta - \alpha'= \underline \beta - \underline \alpha'$
leads to $\alpha + \beta < \underline \alpha + \underline \beta$ and so
to a contradiction. 

Second, let us consider two vectors 
$\bra{\psi(\hat x_1 ,\hat x_2)}$ and 
$\bra{\psi(\hat{\underline{x}}_1, \hat{\underline{x}}_2)}$ 
with $x_1 = \underline{x}_1$ and $x_2 = \underline{x}_2.$
In this case, they are identical: Given a term $\bra{\eta}$ from 
$\bra{\psi(\hat x_1 ,\hat x_2)}$ one can see that this is also
a valid term $\bra{\underline\eta}$ for 
$\bra{\psi(\hat{\underline{x}}_1, \hat{\underline{x}}_2)}$ as follows.  
First, let us assume that 
$\hat{\underline{x}}_1$ contains only one particle in addition to 
$\hat x_1$  (and, consequently, one particle in $\hat x_1$ is missing in
$\hat{\underline{x}}_1$.) Then, one can transform $\hat x_1$ to $\hat{\underline{x}}_1$ by exchanging the two particles, and it is clear
that $\alpha-\beta' = \delta = \underline \alpha - \underline \beta'$
remains valid. By iterating this procedure, one finds that this can
be done for arbitrary $\hat x_1$ and $\hat{\underline{x}}_1$, as long as
$x_1 = \underline{x}_1$. 

Therefore, we can sum the different vectors 
$\ket{\phi(\hat{\underline{x}}_1, \hat{\underline{x}}_2)}$  when 
$x_1 = \underline{x}_1$, and this sum contains
${x_1+x_2 \choose x_1}={x_1+x_2 \choose x_2}$ terms. So, 
the singular value decomposition of $A$
is given by 
\be
A = \sum_{x_1 + x_2 = x} 
\lambda(x_1, x_2)
\ket{\phi(x_1, x_2)}\bra{\psi(x_1 ,x_2)},
\ee
and the singular values are
\bea
\lambda (x_1,x_2) &=& 
\sqrt{
{N-x_1 -x_2 \choose k-x_2}{x_1+x_2 \choose x_2+\delta} 
{N- x_1 - x_2 \choose k-\delta - x_2}{x_1+x_2 \choose x_2}
}/{N \choose k}.
\nonumber
\\
\eea

{\it Step 4.} Finally, we can derive the precise conditions for the matrix 
$X_\delta^{(k)}$ to be positive semidefinite. In our situation, this block 
matrix is positive, iff its Schur complement
\be
Y_\delta^{(k)} = 
\omega_{k-\delta}\mathds{1} - A \frac{1}{\omega_{k+\delta}\mathds{1}} A^T
\ee
is positive semidefinite \cite{positive}. From this it follows, that  $X_\delta^{(k)}$ is 
positive semidefinite iff
\be
\omega_{k-\delta}{\omega_{k+\delta}} \geq \max_{x_1, x_2, x_1+x_2=x}
\{\lambda^2(x_1,x_2)\}.
\ee
Together with the result on the projective entanglement witnesses (Proposition 3), 
this proves Theorem 6.
\hfill $\Box$

\subsection*{A3: Proof of Proposition 9}

In order to show that $\widetilde{\WW}_N$ is an entanglement 
witness, we have to show that 
\begin{equation}
\max_{\psi}\langle\psi|\widetilde{\WW}_N|\psi\rangle\geq 0,
\end{equation}
where the minimization is taken over all biseparable states. Let 
$|\psi\rangle=|a\rangle \otimes|b\rangle$ a pure biseparable state 
with $|a\rangle$ being a state on $K$ qubits and $|b\rangle$ a state 
on $L$ qubits. It is sufficient to restrict ourselves to states with 
one excitation at most, since otherwise the overlap with the $W$ state
vanishes, so that  so that $\langle\psi|\widetilde{\WW}_N|\psi\rangle$
cannot get negative. We can write
\begin{eqnarray}
|a\rangle&=&\alpha|0000\cdots0\rangle+\beta_{1}|100\cdots\rangle+\beta_{2}|010\cdots\rangle+\cdots+\beta_{K}|00\cdots 1\rangle,
\nonumber
\\
|b\rangle&=&\gamma|0000\cdots0\rangle+\delta_{1}|100\cdots\rangle+\delta_{2}|010\cdots\rangle+\cdots+\delta_{L}|00\cdots 1\rangle,
\end{eqnarray}
with complex coefficients $\alpha,\gamma, \beta_{i},\delta_{i}$. In 
the following, we show that the coefficients can be chosen to be real 
and that furthermore $\beta_{i}=\beta_{j}=\beta$ and $\delta_{i}=\delta_{j}=\delta$ 
can be assumed.
To proof this statement, we use the following notation. We denote by 
$|\beta\rangle$ the column vector of the $\beta_{i}$ with $K$ entries 
and $|\delta\rangle$ is the $L$ component vector of 
the $\delta_{i}$. In addition to that, $E_{M\times N}$ denotes a matrix 
with $M$ rows and $N$ columns where all entries are equal to $1$. With 
that notation, one obtains after a short calculation:
\begin{eqnarray}
\langle\psi|\widetilde{W}|\psi\rangle
&=&
\omega_{0}|\alpha|^{2}|\gamma|^{2}+\omega_{1}(|\alpha|^{2}\langle\delta|\delta\rangle+|\gamma|^{2}\langle\beta|\beta\rangle)
+\omega_{2}\langle \delta|\delta\rangle\langle\beta|\beta\rangle
-
\nonumber
\\
&&
-\frac{1}{N}\big(|\alpha|^{2}\langle\delta|E_{L\times L}|\delta\rangle+|\gamma|^{2}\langle\beta|E_{K\times K}|\beta\rangle+
\nonumber
\\
&&
+\alpha^{*}\gamma\langle\delta|E_{L\times K}|\beta\rangle+\alpha\gamma^{*}\langle\beta|E_{K\times L}|\delta\rangle\big).
\end{eqnarray}
Let us fix $|\alpha|$ and $|\gamma|$ which implies that the norm of 
$|\delta\rangle$ and $|\beta\rangle$ is also fixed, because 
$|a\rangle$ and $|b\rangle$ are normalized vectors.
The term $\langle\delta|E_{L\times L}|\delta\rangle$ is maximal,
if $|\delta\rangle$ is the vector corresponding to the maximal 
singular value of $E_{L\times L}$. This is the case iff $\delta_{i}=\delta_{j}=\delta\in\mathds{R}$. For the same 
reason, $\langle\beta|E_{K\times K}|\beta\rangle$ is maximal 
iff $\beta_{i}=\beta_{j}=\beta\in\mathds{R}$. At the same time, 
we maximize $\alpha^{*}\gamma\langle\delta|E_{L\times K}|\beta\rangle+\alpha\gamma^{*}\langle\beta|E_{K\times L}|\delta\rangle=2Re\left(\alpha^{*}\gamma\langle\delta|E_{L\times K}|\beta\rangle\right)$ with this choice of $|\beta\rangle$ and 
$|\delta\rangle$ since the singular vectors of $E_{L\times K}$ 
are of the same type. In addition to that, it is optimal to take 
$\alpha$ and $\gamma$ real  because then
$\alpha^{*}\gamma\langle\delta|E_{L\times K}|\beta\rangle$ is real.

So in the end we have only to perform a minimization over real coefficients $\alpha,\beta,\gamma$ and $\delta$ with the normalization constraints $\alpha^{2}+K\beta^{2}=\gamma^{2}+L\delta^{2}=1$. This leads after a straightforward calculation to the given formula in Proposition 9.
\hfill $\Box$

\section*{References}

\end{document}